\documentclass[10pt,conference]{IEEEtran}
\usepackage{epsfig,setspace,amsmath,epsf,amssymb,bm,theorem,cite,graphicx, epstopdf,algorithm,algpseudocode,float,color,mathtools}
\usepackage[table,xcdraw]{xcolor}

\newtheorem{theorem}{Theorem}

\newtheorem{remark}{Remark}

\IEEEoverridecommandlockouts
\allowdisplaybreaks

\begin{document}

\title{Model Segmentation for Storage Efficient Private Federated Learning with Top $r$ Sparsification}

\author{Sajani Vithana \qquad Sennur Ulukus\\
	\normalsize Department of Electrical and Computer Engineering\\
	\normalsize University of Maryland, College Park, MD 20742 \\
	\normalsize \emph{spallego@umd.edu} \qquad \emph{ulukus@umd.edu}}

\maketitle

\begin{abstract}
In federated learning (FL) with top $r$ sparsification, millions of users collectively train a machine learning (ML) model locally, using their personal data by only communicating the most significant $r$ fraction of updates to reduce the communication cost. It has been shown that the values as well as the indices of these selected (sparse) updates leak information about the users' personal data. In this work, we investigate different methods to carry out user-database communications in FL with top $r$ sparsification efficiently, while guaranteeing information theoretic privacy of users' personal data. These methods incur considerable storage cost. As a solution, we present two schemes with different properties that use MDS coded storage along with a model segmentation mechanism to reduce the storage cost at the expense of a controllable amount of information leakage, to perform private FL with top $r$ sparsification. 

\end{abstract}

\section{Introduction}

Private read-update-write (PRUW) \cite{pruw_jpurnal,ourICC,pruw,sparse,rd,dropout,sparseFL1} is the concept of reading data from and writing updates back to specific sections in a data storage system without revealing the section indices or the values of updates to the data storage. Most applications of PRUW are in distributed learning, specifically in federated learning (FL) \cite{FL1,FL2} where millions of users train various machine learning (ML) models using the private data stored in their local devices. Since each individual user only has access to a limited amount of local data, it is possible that the updates generated by the user for most parameters from the training process are insignificant. Top $r$ sparsification \cite{rtopk,sparse1} is introduced in FL to only upload the most significant $r$ fraction of updates to increase the efficiency of the FL process by reducing communications with insignificant impact. 

However, in top $r$ sparsification, the users send the sparse updates along with their indices, which leak information about each user's private data \cite{comprehensive, featureLeakage, SecretSharer, InvertingGradients,DeepLeakage}. Note that the values as well as the indices of the sparse updates leak information about the user's personal data since the databases are able to find the specific parameters in the model on which the user's data has the most and least impact. In this work, we propose schemes to perform user-database communications in an FL setting with top $r$ sparsification using PRUW to guarantee privacy of the values and the indices of the sparse updates.

\begin{figure}[t]
    \centering
    \includegraphics[scale=0.5]{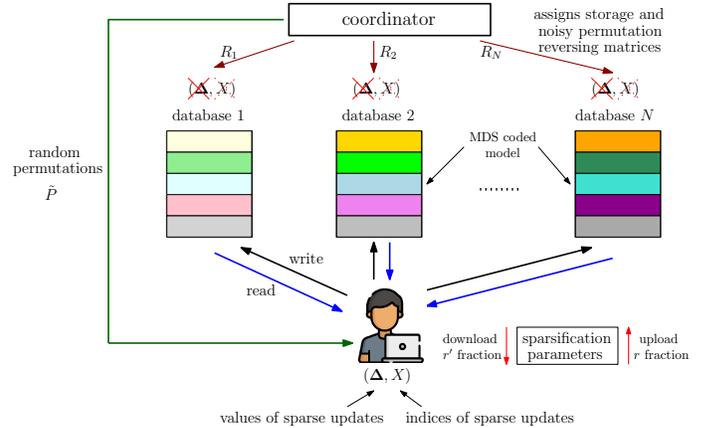}
    \caption{System model.}
    \label{model}
    \vspace*{-0.5cm}
\end{figure}

The system model considered in this work consists of $N$ non-colluding databases storing the MDS coded FL model (Fig.~\ref{model}), that requires the user to download and update the most significant $r'$ and $r$ fractions of subpackets, respectively, without revealing their values or the indices to the databases. In the proposed schemes, we guarantee information theoretic privacy of the values of updates by adding random noise, based on Shannon's one time pad theorem. 

Further, in order to guarantee information theoretic privacy of the indices of the sparse updates, we use a permutation technique where the coordinator in Fig.~\ref{model} initially assigns a random permutation of sets of parameters of the model and makes it available to all users. At the same time, the coordinator places the corresponding \emph{noise added permutation reversing matrices} at the databases. Note that the databases are unaware of the underlying permutation despite having access to the permutation reversing matrices which are noisy, again due to Shannon's one time pad theorem. 

\begin{table*}[ht]
\begin{center}
\begin{tabular}{ |c|c|c|c|c| }
\hline
  case & reading cost & writing cost & storage complexity & information leakage\\ 
  \hline
  Case 1 & $\frac{3r'(1+\frac{\log_q P}{N})}{1-\frac{1}{N}}$ & $\frac{3r(1+\log_q P)}{1-\frac{1}{N}}$ & $O(\frac{L^2}{BN^2})$ & $H(\hat{X}_1,\dotsc,\hat{X}_B)$\\
  \hline
  Case 2 & $\frac{5r'(1+\frac{\log_q P}{N})}{1-\frac{1}{N}}$ & $\frac{5r(1+\log_q P)}{1-\frac{1}{N}}$ & $\max\{O(\frac{L^2}{BN^2}),O(B^2)\}$ & $H(\Tilde{X}_1,\dotsc,\Tilde{X}_B)$\\
  \hline
\end{tabular}
\end{center}
\caption{Achievable sets of communication costs, storage costs and amounts of information leakage.}
\label{main_res}
\vspace*{-0.5cm}
\end{table*}

Once the FL process begins, the coordinator leaves the system, and all communications between the users and databases take place in terms of the permuted indices, which guarantees the privacy of the indices of sparse updates. The permuted updates sent by the users are placed at the correct (non-permuted) positions privately, with the aid of the \emph{noise added permutation reversing matrices} stored at the databases. 

This process incurs a large storage cost due to the large \emph{noise added permutation reversing matrices} as shown in \cite{pruw_jpurnal,sparse}. To alleviate this, we introduce a segmentation mechanism that divides the FL model into $B$ segments, and carries out permutations separately in each segment to hide the indices of the sparse updates. This reduces the storage cost significantly at the expense of a certain amount of information leakage. The amount of information leaked on the indices of the sparse updates can be maintained under a desired privacy leakage budget by varying the number of segments $B$. 

This work differs from \cite{sparseFL1} by using coded storage to achieve lower storage costs at the expense of increased read-write costs. We propose two schemes in this paper to perform private FL with top $r$ sparsification. The first scheme achieves lower read-write costs at the expense of a larger storage cost or information leakage. The second scheme uses an additional round of permutations to reduce the information leakage, at the expense of increased read-write costs. Based on the specifications and limitations of a given FL task, one can choose the most suitable scheme with the optimum number of segments $B$, to perform private FL with top $r$ sparsification.

\section{Problem Formulation}

We consider $N$ non-colluding databases, each storing an FL model consisting of $L$ parameters, which are divided into $P$ subpackets, each containing $\ell=\frac{L}{P}$ parameters. All parameters take values from a large enough finite field $\mathbb{F}_q$. The parameters of each subpacket are combined to obtain a single symbol using an $(\ell,N)$ MDS code, to reduce the storage cost.

Top $r$ sparsification is considered in both uplink and downlink. The process is divided into two phases, namely, the reading phase and the writing phase. In the reading phase at time $t$, each user reads (downloads) a set of $Pr'$ subpackets from all databases, where $r'$, $0\leq r'\leq1$, is the downlink sparsification rate. These $Pr'$ subpackets are determined by the databases, based on the information received by the users in the writing phase of time $t-1$. In the writing phase at time $t$, each user chooses the $Pr$ subpackets with the most significant updates, where $r$, $0\leq r\leq1$, is the uplink sparsification rate, and sends updates corresponding to those $Pr$ subpackets along with their indices to all databases (direct values and indices are not revealed). Note that privacy leakage can occur only in the writing phase since the user does not send any information to databases in the reading phase. The following privacy, security and correctness conditions are considered in this work.

\emph{Privacy of the values of updates:} No information on the values of updates is allowed to leak to the databases, i.e., 
\begin{align}
    I(\Delta_i^{[t]};G_n^{[t]})=0, \quad \forall n, \quad \forall i,
\end{align}
where $\Delta_{i}^{[t]}$ is the $i$th sparse update and $G_{n}^{[t]}$ is all the information sent by the user to database $n$, both at time $t$.

\emph{Privacy of the indices of sparse updates:} The amount of information leaked on the indices of the sparse updates needs to be maintained under a given privacy leakage budget $\epsilon$, i.e.,
\begin{align}
    I(X^{[t]};G_n^{[t]})\leq\epsilon, \quad \forall n, 
\end{align}
where $X^{[t]}$ is the set of indices of the sparse subpackets updated by a given user at time $t$. The system model with the privacy constraints is shown in Fig.~\ref{model}. A coordinator is used to initialize the scheme. 

\emph{Security of the model:} No information about the model parameters is allowed to leak to the databases, i.e.,
\begin{align}
    I(W^{[t]};S_n^{[t]})=0, \quad \forall n,
\end{align}
where $W^{[t]}$ is the FL model and $S_n^{[t]}$ is the data content in database $n$ at time $t$.

\emph{Correctness in the reading phase:} The user should be able to correctly decode the sparse set of subpackets $J$ of the model, from the downloads in the reading phase, i.e.,
\begin{align}
H(W_{J}^{[t-1]}|A_{1:N}^{[t]})=0,
\end{align}
where $W_{J}^{[t-1]}$ subpackets in set $J$ (before updating) and $A_n^{[t]}$ is the information downloaded from database $n$ at time $t$.

\emph{Correctness in the writing phase:} Let $J'$ be the set of most significant $Pr$ subpackets of the model, updated by a user at time $t$. Then, the model should be correctly updated as,
\begin{align}
    W_{s}^{[t]}=
    \begin{cases}
    W_{s}^{[t-1]}+\Delta_{s}^{[t]}, & \text{if $s\in J'$}\\
    W_{s}^{[t-1]}, & \text{if $s\notin J'$}
    \end{cases},
\end{align}
where $W_{s}^{[t-1]}$ is subpacket $s$ of the FL model at time $t-1$, and $\Delta_{s}^{[t]}$ is the corresponding update of subpacket $s$ at time $t$.

\emph{Reading and writing costs:} The reading and writing costs are defined as $C_R=\frac{\mathcal{D}}{L}$ and $C_W=\frac{\mathcal{U}}{L}$, respectively, where $\mathcal{D}$ is the total number of symbols downloaded in the reading phase, $\mathcal{U}$ is the total number of symbols uploaded in the writing phase, and $L$ is the size of the model. 

\emph{Storage complexity:} The storage complexity is quantified by the order of the number of symbols stored in each database. 

In this work, we propose schemes to perform user-database communications in FL with top $r$ sparsification on MDS coded data to reduce the storage cost, and quantify the minimum achievable communication costs while satisfying all privacy, security and correctness conditions described above.

\section{Main Result}

\begin{theorem}
    Consider an FL setting with top $r$ sparsification, where the model with $L$ parameters belonging to $P$ subpackets, each with $\ell$ parameters are stored in $N$ non-colluding databases using an $(\ell,N)$ MDS code. The $P$ subpackets are further divided into $B$ segments, each consisting of the $\frac{P}{B}$ consecutive, non-overlapping subpackets. Let $\hat{X}_i$, $i\in\{1,\dotsc,B\}$ be the random variable representing the number of sparse subpackets updated by a given user from segment $i$, and let $(\Tilde{X}_1,\dotsc,\Tilde{X}_B)$ be the vector representing all distinct combinations of $(\hat{X}_1,\dotsc,\hat{X}_B)$ irrespective of the segment index. Then, the communication costs, storage costs and amounts of information leakage in Table~\ref{main_res} are achievable. 
\end{theorem}

\begin{remark}
    The two cases in Table~\ref{main_res} correspond to the results of two schemes presented in Section~\ref{scheme}. Scheme 1 (case 1) results in a lower communication cost compared to scheme 2, at the expense of a larger information leakage. The information leakage of scheme 2 is smaller than that of scheme 1, i.e., $H(\Tilde{X}_1,\dotsc,\Tilde{X}_B)\leq H(\hat{X}_1,\dotsc,\hat{X}_B)$, since $(\Tilde{X}_1,\dotsc,\Tilde{X}_B)$ combines all different permutations of $(\hat{X}_1,\dotsc,\hat{X}_B)$.  
\end{remark}
    
\begin{remark}
    The information leakage in Table~\ref{main_res} corresponds to the amount of information leaked on the indices of the sparse updates. The number of segments $B$ can be chosen based on the allowed privacy leakage budget $\epsilon$, by solving $H(\hat{X}_1,\dotsc,\hat{X}_B)\leq\epsilon$ or $H(\Tilde{X}_1,\dotsc,\Tilde{X}_B)\leq\epsilon$, based on the chosen case. Information theoretic privacy, i.e., $\epsilon=0$ can be achieved when $B=1$ since $H(\hat{X}_1)=H(\Tilde{X}_1)=H(Pr)=0$.
\end{remark}

\begin{figure}[t]
    \centering
    \includegraphics[scale=0.5]{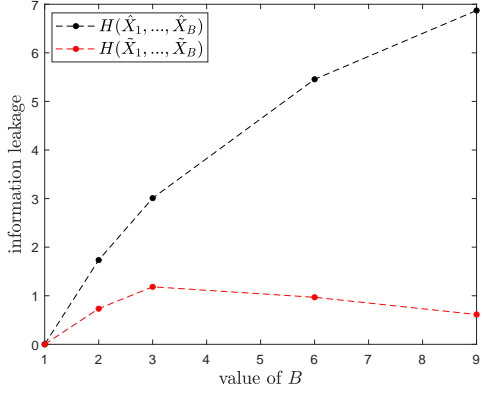}
    \caption{Information leakage of an example setting with $P=18$, $Pr=3$ and different values of $B$.}
    \label{leak}
\end{figure}

\begin{remark}
    Consider an example setting with $P=18$ subpackets divided into $B=1, 2, 3, 6, 9$ segments. Assume that each subpacket is equally probable to be selected to the set of most significant $Pr=3$ subpackets. The behavior of the information leakage for each value of $B$ is shown in Fig.~\ref{leak}. In general, the higher the storage complexity, the lower the information leakage is and vice versa.
\end{remark}

\section{Proposed Schemes}\label{scheme}

\subsection{Case 1: Within-Segment Permutations}

\begin{figure}
		\centering
		\includegraphics[scale=0.55]{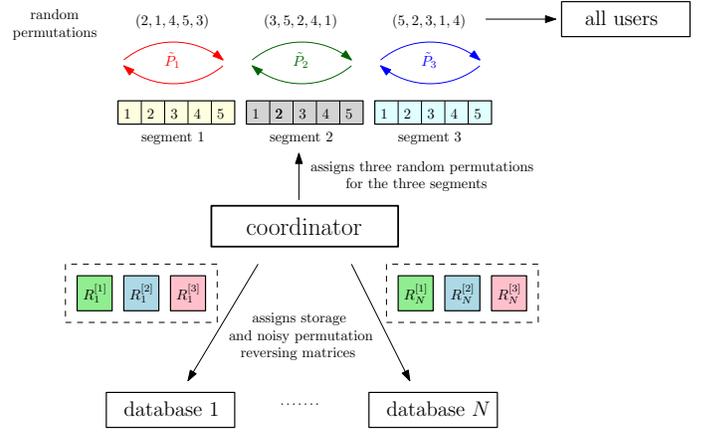}
		\caption{Initialization of the scheme for case 1.}
		\label{init_c1_eg}
		\vspace*{-0.4cm}
\end{figure}

The proposed scheme for case 1 utilizes random noise addition and within-segment permutations to guarantee privacy of the read-write process. The scheme is presented in terms of an example due to limited space here. The example setting is shown in Fig.~\ref{init_c1_eg}, where the FL model consisting of $P=15$ subpackets are grouped into $B=3$ segments, each containing five subpackets. 

\subsubsection{Initialization} 

The storage of a single subpacket $s$ in database $n$, $n\in\{1,\dotsc,N\}$, is given by,
\begin{align}\label{singlesp}
    S_n^{[s]}=\sum_{i=1}^\ell \frac{1}{\alpha_n^i}W_i^{[s]}+\sum_{i=0}^x \alpha_n^i Z_{s,i},
\end{align}
with $x=\ell$, where $W_i^{[s]}$ is the $i$th parameter of subpacket $s$, $Z_{s,i}$ is a random noise symbol, $\ell$ is the size of a subpacket (subpacketization) and $\alpha_n$s are globally known distinct constants from $\mathbb{F}_q$. Therefore, the storage of segment $j$, $j\in\{1,2,3\}$, each consisting of five subpackets is given by,
 \begin{align}\label{storage_c2_eg}
 S_{n,j}=\begin{bmatrix}
    \sum_{i=1}^\ell \frac{1}{\alpha_n^i}W_i^{[1,j]}+\sum_{i=0}^x \alpha_n^i Z_{1,i}\\
    \vdots\\
    \sum_{i=1}^\ell \frac{1}{\alpha_n^i}W_i^{[\frac{P}{B},j]}+\sum_{i=0}^x \alpha_n^i Z_{\frac{P}{B},i}
    \end{bmatrix},
\end{align}
with $x=\ell$ and $\frac{P}{B}=5$, where $W_i^{[k,j]}$ is the $i$th symbol of subpacket $k$ of segment $j$. Before the FL process begins, the coordinator randomly chooses three permutations for the five subpackets in each of the $B=3$ segments, from the $5!$ options. Let the permutations assigned for the three segments be $\Tilde{P}_1=\{2,1,4,5,3\}$, $\Tilde{P}_2=\{3,5,2,4,1\}$ and $\Tilde{P}_3=\{5,2,3,1,4\}$, respectively. This is known by all participating users, but not the databases. The coordinator also places the corresponding three noise added permutation reversing matrices at each of the databases as shown in Fig.~\ref{init_c1_eg}. For example, the noise added permutation reversing matrix corresponding to the first segment, stored at database $n$, $n\in\{1,\dotsc,N\}$, is given by,
\begin{align}\label{R1_c2}
    R_n^{[1]}=\begin{bmatrix}
        0 & 1 & 0 & 0 & 0\\
        1 & 0 & 0 & 0 & 0\\
        0 & 0 & 0 & 0 & 1\\
        0 & 0 & 1 & 0 & 0\\
        0  & 0 & 0 & 1 & 0\\
    \end{bmatrix}+\alpha_n^\ell\bar{Z}_1,
\end{align}
where $\bar{Z}_1$ is a matrix of size $5\times5$, consisting of random elements from $\mathbb{F}_q.$ Note that the binary matrix in \eqref{R1_c2} reverses the permutation $\Tilde{P}_1$, while the noise component $\alpha_n^\ell\bar{Z}_1$ ensures that the databases learn nothing about the underlying permutation from the noise added permutation reversing matrices.

\subsubsection{Reading Phase} 

In the reading phase, the databases decide on a set of $Pr'$ sparse subpackets to be sent to the users at time $t$, based on the sparse updates received at time $t-1$ (for example, the most commonly updated $Pr'$ subpackets). Note that all communications between users and databases take place in terms of the permuted indices of subpackets. Therefore, the $Pr'$ sparse subpackets selected to be sent to the users are also indicated by their permuted indices. Let $\tilde{V}_j$ be the set of permuted indices of the sparse subpackets chosen from segment $j$ to be sent to the users for $j\in\{1,2,3\}$. For example, let $\Tilde{V}_1=\{1,3\}$ be the permuted set of sparse subpackets of segment 1 that needs to be sent to the users at time $t$. One designated database sends the permuted subpacket indices of each segment (segment 1: $\Tilde{V}_1=\{1,3\}$) to the users, from which the users identify the corresponding real sparse subpacket indices using the known permutations. For example, the real indices $V_1$ corresponding to $\Tilde{V}_1=\{1,3\}$ are given by $V_1=\{\Tilde{P}_1(1),\Tilde{P}_1(3)\}=\{2,4\}$. 

Once the permuted indices of the sparse subpackets are sent to the users, each database generates a query to send each sparse subpacket. The query corresponding to the $i$th permuted sparse subpacket of segment $j$, i.e., $\Tilde{V}_j(i)$, is given by $Q_n^{[\Tilde{V}_j(i)]}=R_n^{[j]}(:,\Tilde{V}_j(i))$ for database $n$, $n\in\{1,\dotsc,N\}$. For example, the query corresponding to the first permuted subpacket of segment 1, i.e., $\Tilde{V}_1(1)=1$, is 
\begin{align}\label{q_1}
    Q_n^{[\Tilde{V}_1(1)]}&=Q_n^{[1]}=R_n^{[1]}(:,1)=
    [0,1,0,0,0]^T+\alpha_n^\ell\hat{Z}_1, 
\end{align}
where $\hat{Z}_1$ is the first column of $\bar{Z}_1$ in \eqref{R1_c2}. Then, database $n$, $n\in\{1,\dotsc,N\}$, sends the answer to the query in \eqref{q_1} as
\begin{align}    A_n^{[\tilde{V}_1(1)]}&=S_{n,1}Q_n^{[\tilde{V}_1(1)]}
    =\sum_{i=1}^\ell\frac{1}{\alpha_n^i}W_i^{[2,1]}+P_{\alpha_n}(2\ell),\label{ans2}
\end{align}
where $P_{\alpha_n}(2\ell)$ is a polynomial in $\alpha_n$ of degree $2\ell$. The users then obtain the parameters of the real subpacket 2, i.e., $V_1(1)=\Tilde{P}_1(\Tilde{V}_1(1))=2$, by solving
\begin{align}\label{mat_c2}
    \begin{bmatrix}        \!A_1^{\Tilde{V}_1(1)}\!\\\!\vdots\!\\\!A_N^{\Tilde{V}_1(1)}\!
    \end{bmatrix}
\!=\!
& \begin{bmatrix}
    \frac{1}{\alpha_1^\ell}\!\! & \!\!\dotsc \!\!&\!\! \frac{1}{\alpha_1} \!\!&\!\! 1 \!\!&\!\! \alpha_1 \!&\! \dotsc \!\!&\!\! \alpha_1^{2\ell}\\
    \vdots \!\!&\!\! \vdots \!\!&\!\! \vdots \!\!&\!\! \vdots \!\!&\!\! \vdots \!\!&\!\! \vdots \!\!&\!\! \vdots\\
    \frac{1}{\alpha_N^\ell} \!\!&\!\! \dotsc \!\!&\!\! \frac{1}{\alpha_N} \!\!&\!\! 1 \!\!&\!\! \alpha_N \!\!\!&\!\!\! \dotsc \!\!&\!\! \alpha_N^{2\ell}\\
\end{bmatrix}\!\!
\begin{bmatrix}
        W_{\ell}^{[2,1]}\\\vdots\\W_{1}^{[2,1]}\\ R_{0:2\ell}
    \end{bmatrix}
\end{align}
where $R_i$ are the coefficients of $P_{\alpha_n}(2\ell)$. Note that \eqref{mat_c2} is solvable given that $N=3\ell+1$, which determines the subpacketization as $\ell=\frac{N-1}{3}$. The same procedure described above is carried out for all sparse subpackets in each of the $B=3$ segments. The resulting reading cost is given by,
\begin{align}
    C_R&=\frac{Pr'\log_q P+Pr'N}{L}=\frac{3r'(1+\frac{\log_q P}{N})}{1-\frac{1}{N}}.
\end{align}

\subsubsection{Writing Phase}

After carrying out the training process locally, the user chooses the $Pr$ subpackets with the most significant set of updates, and sends combined updates corresponding to each of the selected $Pr$ subpackets, along with their permuted indices. The combined update of the $i$th subpacket of segment $j$ is defined, assuming this subpacket is among the sparse set, as
\begin{align}\label{comb}
    U_n^{[i,j]}=\sum_{k=1}^\ell \frac{1}{\alpha_n^k}\Delta_{k}^{[i,j]}+Z^{[i,j]},
\end{align}
where $\Delta_{k}^{[i,j]}$ is the update of the $k$th symbol of the $i$th subpacket of segment $j$ and $Z^{[i,j]}$ is a random noise symbol. 

For example, assume that the user chooses to send the updates of real subpackets 2 and 4 from segment 1, subpacket 2 from segment 2 and subpacket 5 from segment 3. Note that the permuted subpacket index corresponding to the real subpacket 2 of segment 1 is 1, based on $\Tilde{P}_1=\{2,1,4,5,3\}$. Therefore, the permuted\footnote{For case 1, we only consider permutations within segments, and not among segments. Therefore, the real segment index is revealed to the databases.} (update, subpacket, segment) tuple corresponding to the first sparse update (real subpacket 2 of segment 1), which is sent by the user to database $n$, $n\in\{1,\dotsc,N\}$, is $(U_n^{[2,1]},1,1)$. Similarly, the rest of the permuted (update, subpacket, segment) tuples for this example are given by $(U_n^{[4,1]},3,1)$, $(U_n^{[2,2]},3,2)$ and $(U_n^{[5,3]},1,3)$, based on the rest of the initial permutations, $\Tilde{P}_2=\{3,5,2,4,1\}$ and $\Tilde{P}_3=\{5,2,3,1,4\}$. Each of these permuted tuples are sent to database $n$, $n\in\{1,\dotsc,N\}$, by the user. Once the databases receive the $Pr$ permuted (update, subpacket, segment) tuples, they create the permuted update vectors $Y_n^{[j]}$ for each segment $j$, $j\in\{1,2,3\}$. For the example considered, the permuted update vectors of the three segments are given by,
\begin{align}
    Y_n^{[1]}&=[U_n^{[2,1]},0,U_n^{[4,1]},0,0]^T\\
    Y_n^{[2]}&=[0,0,U_n^{[2,2]},0,0]^T\\
    Y_n^{[3]}&=[U_n^{[5,3]},0,0,0,0]^T,
\end{align}
for database $n$. Using these permuted update vectors and the noise added permutation reversing matrices stored, database $n$, $n\in\{1,\dotsc,N\}$, privately rearranges the updates in the correct order as $\bar{U}_n^{[j]}=R_n^{[j]}Y_n^{[j]}$, $j\in\{1,2,3\}$. For example, the privately rearranged update vector in the correct order for segment 1 in database $n$ is given by,
\begin{align}    
    \bar{U}_n^{[1]}&\!=\!R_n^{[1]}Y_n^{[1]}\!=\!\!\left(\!\begin{bmatrix}
        0 & 1 & 0 & 0 & 0\\
        1 & 0 & 0 & 0 & 0\\
        0 & 0 & 0 & 0 & 1\\
        0 & 0 & 1 & 0 & 0\\
        0  & 0 & 0 & 1 & 0\\    \end{bmatrix}\!+\!\alpha_n^\ell\bar{Z}_1\!\!\right)
        \!\!\!\begin{bmatrix}U_n^{[2,1]}\\0\\U_n^{[4,1]}\\0\\0
    \end{bmatrix}\\
    &=\!\!\begin{bmatrix}
        0\\U_n^{[2,1]}\\0\\U_n^{[4,1]}\\0
    \end{bmatrix}\!+\!P_{\alpha_n}(\ell)\!=\!\begin{bmatrix}
        0\\\sum_{i=1}^\ell \frac{1}{\alpha_n^i}\Delta_{i}^{[2,1]}\\0\\\sum_{i=1}^\ell \frac{1}{\alpha_n^i}\Delta_{i}^{[4,1]}\\0
    \end{bmatrix}\!+\!P_{\alpha_n}(\ell), \label{incr_c2}
\end{align}
where $P_{\alpha_n}(\ell)$ here is a vector of size $5\times1$, consisting of polynomials in $\alpha_n$ of degree $\ell$. Note that the  updates of real subpackets 2 and 4 in segment 1 are now placed correctly in \eqref{incr_c2} at the $2$nd and $4$th positions, without the knowledge of the databases. Since the incremental update of each segment \eqref{incr_c2} is in the same form as the storage in \eqref{storage_c2_eg}, the incremental update is directly added to the existing storage to obtain the updated version, i.e., $S_{n,j}(t)=S_{n,j}(t-1)+\bar{U}_n^{[j]}$, $j\in\{1,2,3\}$ in each database. The writing cost for case 2 is given by,
\begin{align}
    \!C_W=\frac{PrN(1+\log_q B+\log_q \frac{P}{B})}{L}=\frac{3r(1+\log_qP)}{1-\frac{1}{N}}.
\end{align}
The total storage complexity is given by 
\begin{align}
O(P)+O\left(\frac{P^2}{B^2}\times B\right)=O\left(\frac{P^2}{B}\right)=O\left(\frac{L^2}{BN^2}\right).
\end{align}

\subsection{Case 2: Within-Segment and Inter-Segment Permutations}

\begin{figure}
		\centering
		\includegraphics[scale=0.5]{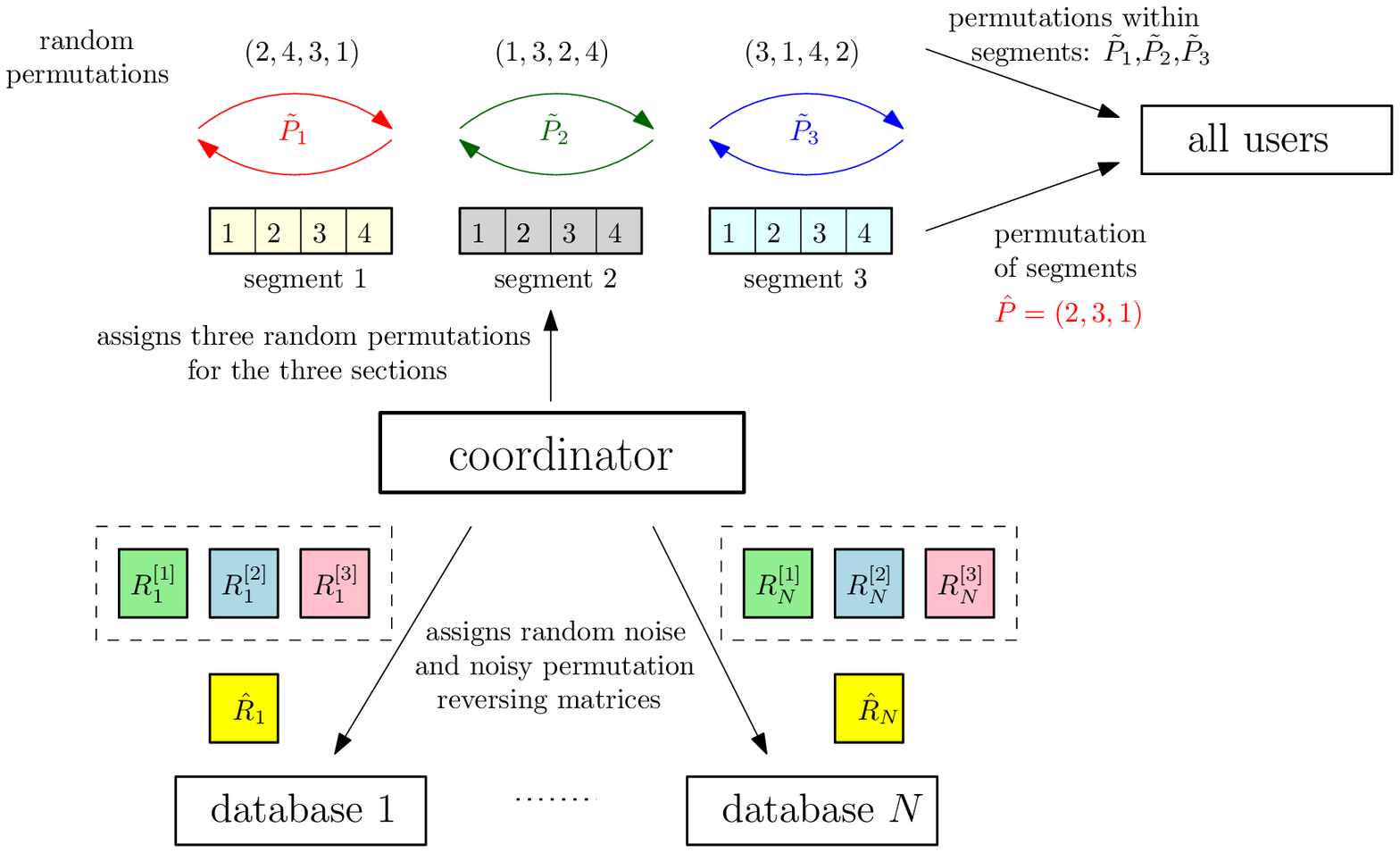}
		\caption{Initialization of the scheme for case 2.}
		\label{init_c3_eg}
\end{figure}

In addition to noise addition and within-segment permutations considered in case 1 to guarantee the required privacy constraints, we consider inter-segment permutations as well in case 2 to achieve higher privacy guarantees. The scheme is presented in terms of an example, which is shown in Fig.~\ref{init_c3_eg}, where there are $12$ subpackets, divided into three segments.

\subsubsection{Initialization}

The storage of a single subpacket $s$ is the same as \eqref{singlesp} with $x=2\ell$, and the storage of a given segment $j$, $j\in\{1,2,3\}$, is the same as \eqref{storage_c2_eg} with $x=2\ell$ and $\frac{P}{B}=4$. As described in case 1, the coordinator randomly chooses the three within-segment permutations $\Tilde{P}_1$, $\Tilde{P}_2$ $\Tilde{P}_3$ and the inter-segment permutation $\hat{P}$, and sends them to the users as shown in Fig.~\ref{init_c3_eg}. The coordinator also places the corresponding four noise added permutation reversing matrices given by $R_n^{[1]}$,$R_n^{[2]}$, $R_n^{[3]}$ and $\hat{R}_n$ at database $n$, $n\in\{1,\dots,N\}$. For instance, the noise added permutation reversing matrix corresponding to the first within-segment permutation $\tilde{P}_1=\{2,4,3,1\}$ in the example considered in Fig.~\ref{init_c3_eg} is given by,
\begin{align}
    R_n^{[1]}=\begin{bmatrix}
        0 & 0 & 0 & 1\\
        1 & 0 & 0 & 0\\
        0 & 0 & 1 & 0\\
        0 & 1 & 0 & 0
    \end{bmatrix}+\alpha_n^{\ell}\bar{Z}_1,
\end{align}
where $\Bar{Z}_1$ is a random noise matrix of size $4\times4$. The noise added permutation reversing matrix corresponding to the inter-segment permutation $\hat{P}$ is given by,
\begin{align}
\hat{R}_n=\begin{bmatrix}
        0 & 0 & 1\\
        1 & 0 & 0\\
        0 & 1 & 0
\end{bmatrix}+\alpha_n^{\ell} Z,
\end{align}
where $Z$ is a random noise matrix of size $3\times3$. To aid the calculations of this scheme, we combine the two types of noise added permutation reversing matrices to obtain a combined noisy permutation reversing matrix (this is not stored at databases). For the example considered, the combined noisy permutation reversing matrix of database $n$ is given by,
\begin{align}
    R_n&=\begin{bmatrix}
        R_n^{[1]} & 0_{4\times4} & 0_{4\times4}\\
        0_{4\times4} & R_n^{[2]} & 0_{4\times4}\\
        0_{4\times4} & 0_{4\times4} & R_n^{[3]}
    \end{bmatrix}\times (\hat{R}_n\otimes I_4)\\ 
    &=\begin{bmatrix}
        0_{4\times4} & 0_{4\times4} & \begin{bmatrix}
        0 & 0 & 0 & 1\\
        1 & 0 & 0 & 0\\
        0 & 0 & 1 & 0\\
        0 & 1 & 0 & 0
    \end{bmatrix}\\
    \begin{bmatrix}
        1 & 0 & 0 & 0\\
        0 & 0 & 1 & 0\\
        0 & 1 & 0 & 0\\
        0 & 0 & 0 & 1
    \end{bmatrix} & 0_{4\times4} & 0_{4\times4}\\
    0_{4\times4} & \begin{bmatrix}
        0 & 1 & 0 & 0\\
        0 & 0 & 0 & 1\\
        1 & 0 & 0 & 0\\
        0 & 0 & 1 & 0    
        \end{bmatrix} & 0_{4\times4}
    \end{bmatrix}\nonumber\\
    &\quad +\alpha_n^{\ell}P_{\alpha_n}(\ell),
\end{align}
where $I_4$ is the identity matrix of size $4\times4$ and $P_{\alpha_n}(\ell)$ here is a matrix of size $12\times12$ with entries consisting of polynomials of $\alpha_n$ of up to degree $\ell$. 

\subsubsection{Reading Phase}

In the reading phase, the databases determine the set of $Pr'$ subpackets to be sent to the users at time $t$, based on the permuted information received by all users in the writing phase of time $t-1$, as explained in case 1. Since both subpacket and segment indices received by the users are in terms of their permuted indices, the $Pr'$ subpackets chosen by the databases in the reading phase are also indicated by their permuted indices. Let the permuted (subpacket, segment) tuples of the $Pr'$ subpackets to be sent to the users be denoted by $(\eta_p,\phi_p)$. This information is sent to all users by one designated database. For example, assume that the designated database sends the permuted (subpacket, segment) tuples given by $(\eta_p,\phi_p)=\{(1,3), (1,1), (1,2)\}$. These permuted tuples can be converted to their real indices using the permutations known by the users as follows. Consider the first permuted pair $(1,3)$. Since the permuted segment index is $\phi_p=3$, the corresponding real segment index is $\phi_r=\hat{P}(3)=1$. Then, the user can decode the subpacket index within the first segment as, $\eta_r=\Tilde{P}_1(1)=2$. Therefore, the real (subpacket, segment) pair corresponding to the permuted (subpacket,segment) pair $(\eta_p,\phi_p)=(1,3)$ is given by $(\eta_r,\phi_r)=(2,1)$. Similarly, the real set of sparse subpacket indices corresponding to the three permuted pairs are given by $ (\eta_r,\phi_r)=\{(2,1),(1,2),(3,3)\}$.

In order to send the subpacket corresponding to $(\eta_p,\phi_p)=(i,j)$, database $n$, $n\in\{1,\dots,N\}$, creates a query given by,
\begin{align}
    Q_n^{[i,j]}&=R_n(:,(j-1)\frac{P}{B}+i).
\end{align}
For $(\eta_p,\phi_p)=(1,3)$, the corresponding query is given by,
\begin{align}\label{queg1}
    Q_n^{[1,3]}&=R_n(:,9)=[[0,1,0,0],0_{1\times8}]^T
+\alpha_n^{\ell}P_{\alpha_n}(\ell),
\end{align}
where $P_{\alpha_n}(\ell)$ is a vector of size $12\times1$ with entries consisting of polynomials of $\alpha_n$ of degrees up to $\ell$. The corresponding answer to the query in \eqref{queg1} is given by,
\begin{align}
    A_n^{[1,3]}&=S_n^TQ_n^{[1,3]}=\sum_{i=1}^\ell \frac{1}{\alpha_n^i}W_i^{[2,1]}+P_{\alpha_n}(4\ell).   
\end{align}
The users can obtain the parameters of the second subpacket of segment 1 (since the real indices corresponding to permuted $(1,3)$ are $(2,1)$) using the $N$ answers received if $N=5\ell+1$ is satisfied. This defines the subpacketization for case 2 as $\ell=\frac{N-1}{5}$. The resulting reading cost is given by,
\begin{align}
    C_R&=\frac{Pr'(N+\log_q B+\log_q\frac{P}{B})}{L}=\frac{5r'(1+\frac{\log_q P}{N})}{1-\frac{1}{N}}.
\end{align}

\subsubsection{Writing Phase}

In the writing phase, each user selects the $Pr$ subpackets with the most significant updates, and sends the corresponding combined updates along with their permuted subpacket and segment indices to all databases. Let $(\eta_r^{[m]},\phi_r^{[m]})$, $m\in\{1,\dotsc,Pr\}$, be the real (subpacket, segment) information of the $m$th sparse subpacket. The combined update of the $m$th sparse subpacket is given by \eqref{comb} with $i=\eta_r^{[m]}$ and $j=\phi_r^{[m]}$. For the example considered in Fig.~\ref{init_c3_eg}, assume that a user wants to update real (subpacket, segment) pairs given by $(\eta_r,\phi_r)=\{(2,1),(2,2),(3,3)\}$. Based on the within- and inter-segment permutations given by $\Tilde{P}_1=(2,4,3,1)$, $\Tilde{P}_2=(1,3,2,4)$, $\Tilde{P}_3=(3,1,4,2)$ and $\hat{P}=(2,3,1)$, the user sends the permuted (update, subpacket, segment) tuples corresponding to each of the $Pr$ subpackets to all databases. Consider the first sparse subpacket denoted by $(\eta_r,\phi_r)=(2,1)$. The permuted subpacket index corresponding to $\eta_r=2$ when $\phi_r=1$ is given by $\eta_p=\Tilde{P}_{\phi_r}^{-1}(2)=1$. The permuted segment index corresponding to $\phi_r=1$ is given by $\phi_p=\hat{P}^{-1}(1)=3$. Therefore, the permuted (update, subpacket, segment) tuple corresponding to the first sparse subpacket, sent to database $n$ is given by $(U_n^{[2,1]},1,3)$. Similarly, the tuples corresponding to the other sparse subpackets are given by $(U_n^{[2,2]},3,1)$ and $(U_n^{[3,3]},1,2)$. Similar to case 1, the databases create the permuted update vector based on the permuted tuples received by the user. For the example considered, the permuted update vector is given by,
\begin{align}
    Y_n=[0,0,U_n^{[2,2]},0,U_n^{[3,3]},0,0,0,U_n^{[2,1]},0,0,0]^T.
\end{align}
Then, each database calculates the permutation-reversed incremental update as,
\begin{align}
    \bar{U}_n&=R_nY_n=\left[0,\sum_{i=1}^\ell \frac{1}{\alpha_n^i}\Delta_{i}^{[2,1]},0,0,0,\sum_{i=1}^\ell \frac{1}{\alpha_n^i}\Delta_{i}^{[2,2]},0,0,\right.\nonumber\\
    &\qquad\qquad\qquad\left. 0,0,\sum_{i=1}^\ell \frac{1}{\alpha_n^i}\Delta_{i}^{[3,3]},0\right]^T+P_{\alpha_n}(2\ell), \label{last_c4}
\end{align}
where $P_{\alpha_n}(2\ell)$ is a vector of size $12\times1$, consisting of polynomials of $\alpha_n$ of degrees up to $2\ell$. Note that the (real) subpacket 2 of segment 1, subpacket 2 of segment 2 and subpacket 3 of segment 3 ($(\eta_r,\phi_r)=\{(2,1),(2,2),(3,3)\}$) are correctly updated in \eqref{last_c4}, without revealing the real indices to the databases. Since the incremental update in \eqref{last_c4} is in the same form as \eqref{storage_c2_eg} with $x=2\ell$ and $\frac{P}{B}=4$, it is directly added to the existing storage to obtain the updated storage. The writing cost is given by,
\begin{align}
    \!C_W=\frac{PrN(1+\log_q B+\log_q \frac{P}{B})}{L}=\frac{5r(1+\log_q P)}{1-\frac{1}{N}}.
\end{align}
The storage complexities of data, noise added within-segment and inter-segment permutation reversing matrices are given by $O(P)=O(\frac{L}{N})$, $O(\frac{P^2}{B})=O(\frac{L^2}{N^2B})$ and $O(B^2)$, respectively. Therefore, the storage complexity is $\max\{O(\frac{L^2}{N^2B}),O(B^2)\}$.

The proofs of the expressions for the amounts of information leaked on the sparse update indices for arbitrary $B$ (stated in in Table~\ref{main_res}) for cases 1 and 2 are omitted due to limited space.

\bibliographystyle{unsrt}
\bibliography{referencesCISS}

\end{document}